\RequirePackage{fixltx2e}
\documentclass[8.5pt,twoside,twocolumn]{article}
\usepackage{graphicx,subfigure,framed} 

\begin{document}

\twocolumn[
  \begin{@twocolumnfalse}
\noindent\LARGE{\textbf{Al\textsubscript{20}\textsuperscript{+} does melt, albeit above the bulk melting temperature of aluminium.}}
\vspace{0.6cm}

\center\noindent\large{\textbf{Udbhav Ojha,\textit{$^{a}$} Krista G. Steenbergen,\textit{$^{b}$} and
Nicola Gaston$^{\ast}$\textit{$^{a}$}}}\vspace{0.5cm}

{\textit{$^{a}$~The MacDiarmid Institute for Advanced Materials and Nanotechnology, Victoria University of Wellington, P.O. Box 600, Wellington 6012, New Zealand. E-mail: nicola.gaston@vuw.ac.nz}}
{\textit{$^{b}$~Department of Chemistry, University of Kansas, Lawrence, Kansas 66045, USA}}
\newline

\noindent \normalsize{Employing first principles parallel tempering molecular dynamics in the microcanonical ensemble, we report the presence of a clear solid-liquid-like melting transition in Al\textsubscript{20}\textsuperscript{+} clusters, not found in experiments. The phase transition temperature obtained from the multiple histogram method is 993 K, 60 K above the melting point of aluminium. Root mean squared bond length fluctuation, the velocity auto-correlation function and the corresponding power spectrum further confirm the phase transition from a solid-like to liquid-like phase. Atoms-In-Molecules analysis shows a strong charge segregation between the internal and surface atoms, with negatively charged internal atoms and positive charge at the surface. Analysis of the calculated diffusion coefficients indicates different mobilities of the internal and surface atoms in the solid-like phase, and the differences between the environment of the internal atoms in these clusters with that of the bulk atoms suggest a physical picture for the origin of greater-than-bulk melting temperatures.}
\vspace{0.5cm}
 \end{@twocolumnfalse}
  ]

\section{Introduction}
Atomic clusters, viewed as the basic building blocks of nanoscience, have been extensively studied in the past few decades\cite{baletto}. In the size range where surfaces effects cannot be neglected, finite size effects can impart properties to clusters that are significantly different from their bulk counterparts. A preference for noncrystalline global minimum structures\cite{thirteen_atom}, differences in the electronic structure and bonding\cite{mg_bond, bond_LiAl} and changes in thermodynamic behaviour\cite{haberland_sodium, jarrold_GaJACS,jarrold_PRB} have been reported. Not only do they help in understanding the complexities arising at the nanoscale, tunability of the associated intrinsic properties have raised hopes for potential future applications\cite{haberland,martin,alivisatos,andres}.  

The solid to liquid phase transition has been a topic of great interest in these finite systems and significant efforts have gone into developing the corresponding theory\cite{takagi, buffat-borel,jesser,tin,briant_argon,berry,labastie,wales}. Owing to their finite sizes, changes occur rather gradually, spread across a range of temperatures called `phase change' regions\cite{Neirotti}. Pawlow predicted a depression in melting temperature with size\cite{pawlow} but exceptions to this behaviour have already been reported for Ga\cite{jarrold_GaPRL} and Sn\cite{jarrold_tin} clusters. Jarrold and coworkers\cite{jarrold_GaJACS} found specific sizes of gallium clusters for which the specific heat curves were sharply peaked and the addition or subtraction of one atom led to significant changes in the melting behaviour. Furthermore, the absence of any specific heat peak raises questions about the nature of melting in some clusters\cite{jarrold_PRB} while multiple peaks in the specific heat, indicating the observation of multiple stages in cluster melting,\cite{jarrold_twopeaks} have also been reported. All these different observations regarding the phase changes in atomic clusters reflects the sensitivity of the thermodynamical properties to the underlying geometric and electronic structure of the clusters\cite{jarrold_rev}.

Among atomic clusters, aluminium has received considerable attention, both theoretically and experimentally. In the size range of 16-128 atoms, the cationic aluminium clusters show drastic variations in the melting behaviour, including single, multiple, or absent peaks\cite{jarrold_rev}. High temperature superconductivity applications of anionic aluminium clusters have also been explored\cite{al_HTS}. 
  
Experimental specific heat curves have been reported for sizes as low as 16 atoms for Al cations, but the presence of a peak in specific heat curves is limited to sizes above 28 atoms\cite{Al16-48}. Thus cluster melting either happens over a range of temperatures at these sizes, or, alternatively, the observed temperature window for cluster sizes below 28 atoms is insufficient for the melting transition to have been observed. To our knowledge, no first principles-based molecular dynamics study has been done to test this explanation. We employ density functional theory (DFT), found to successfully capture the phase changes in atomic clusters, to study the solid-liquid-like phase transition in Al\textsubscript{20}\textsuperscript{+} cluster. Our aim is not only to capture the melting-like transition but also to explore the dynamics of this finite system. The organization of this work is as follows: Section 2 describes the computational method adopted to study the melting-like transition of Al\textsubscript{20}\textsuperscript{+} cluster. Section 3 presents the results obtained employing various statistical tools and indices to understand the spatial and temporal changes occurring during the phase transition. In section 4 we have compared the electronic structure and overall structural mobility of Al\textsubscript{20}\textsuperscript{+} and Ga\textsubscript{20}\textsuperscript{+} clusters in order to examine the differences between these two isoelectronic yet distinct group 13 elements. Finally, we summarize and conclude in section 5.
 
\section{Computational Details}
Plane wave-based DFT calculations as implemented in the Vienna \textit{ab-initio} simulation package (VASP)\cite{vasp1,vasp2,vasp3,vasp4} were carried out using the projector-augmented wave (PAW)\cite{paw1,paw2} method and the generalized gradient approximation in the Perdew-Wang form (GGA-PW91)\cite{PW91a,PW91b}. An energy cut-off of 350 eV was used during the geometry optimization, which was reset to the default value for molecular dynamics (MD) runs. Only the $\Gamma$-point was used to sample the Brillouin zone. The stacked plane (SP) configuration\cite{krista1,self} identified as the global minimum Ga\textsubscript{20} structure\cite{Ga20structure}, an assembly of alternating hexagonal and pentagonal planes separated by a single atom as shown in Fig. \ref{structure}(a), served as the starting structure for Al\textsubscript{20}\textsuperscript{+} calculations.

The VASP implementation of Born-Oppenheimer molecular dynamics (BOMD) was used to carry out the molecular dynamics calculations. We chose 29 different temperatures between 250 K and 1650 K with 50 K intervals to capture the solid-liquid-like phase transition. In order to observe the higher-temperature dynamics, the temperature range selected is above the temperatures scanned for capturing the melting transition in experiments\cite{Al16-48}. Canonical equilibrations at each temperature were used to gauge initial velocities, and thus set the temperatures to the range of interest for subsequent microcanonical simulations, and so were not used in the analyses. Each of the 29 microcanonical trajectories were run for 47.3 ps with a step size of 1 fs. Compared to our previous MD study on gallium clusters\cite{krista1} a smaller time step was used to counter the energy increase occurring during the microcanonical runs. 

In order to minimise dependence of the thermodynamic quantities on the choice of starting structure, as well as the possibility of the cluster being trapped in one of the local minima in the potential energy surface (PES) of Al\textsubscript{20}\textsuperscript{+}, parallel tempering molecular dynamics (PTMD) was used. The 29 microcanonical runs, each representing a different trajectory, were performed for 100 time steps after which two configurations were randomly selected. Swapping between these two selected configurations was done on a Monte Carlo-based acceptance criteria\cite{CalvoPTMD}. 

Calculation of continuous thermodynamic quantities in the canonical and microcanonical ensemble was done using the multiple histogram (MH) method\cite{CalvoPTMD, krista1}. Configurational and total density of states, which yielded the configurational and total entropies, were calculated using the 29 different potential energy distributions. The entropies led to the calculation of specific heats and temperatures in the desired ensembles. A sliding window analysis technique spanning over ten consecutive time windows was used as a measure of relative convergence of the canonical and microcanonical specific heat curves\cite{krista1}. The Quantum Theory of Atoms-In-Molecules (QTAIM) was used for the spatial charge distribution analysis using freely-available software\cite{henkelman}.   

\section{Results}
\subsection{Phase transitions}
\begin{figure}[t]
\centering
\includegraphics[trim = 0cm 0cm 0cm 0cm,clip=true,width=\columnwidth]{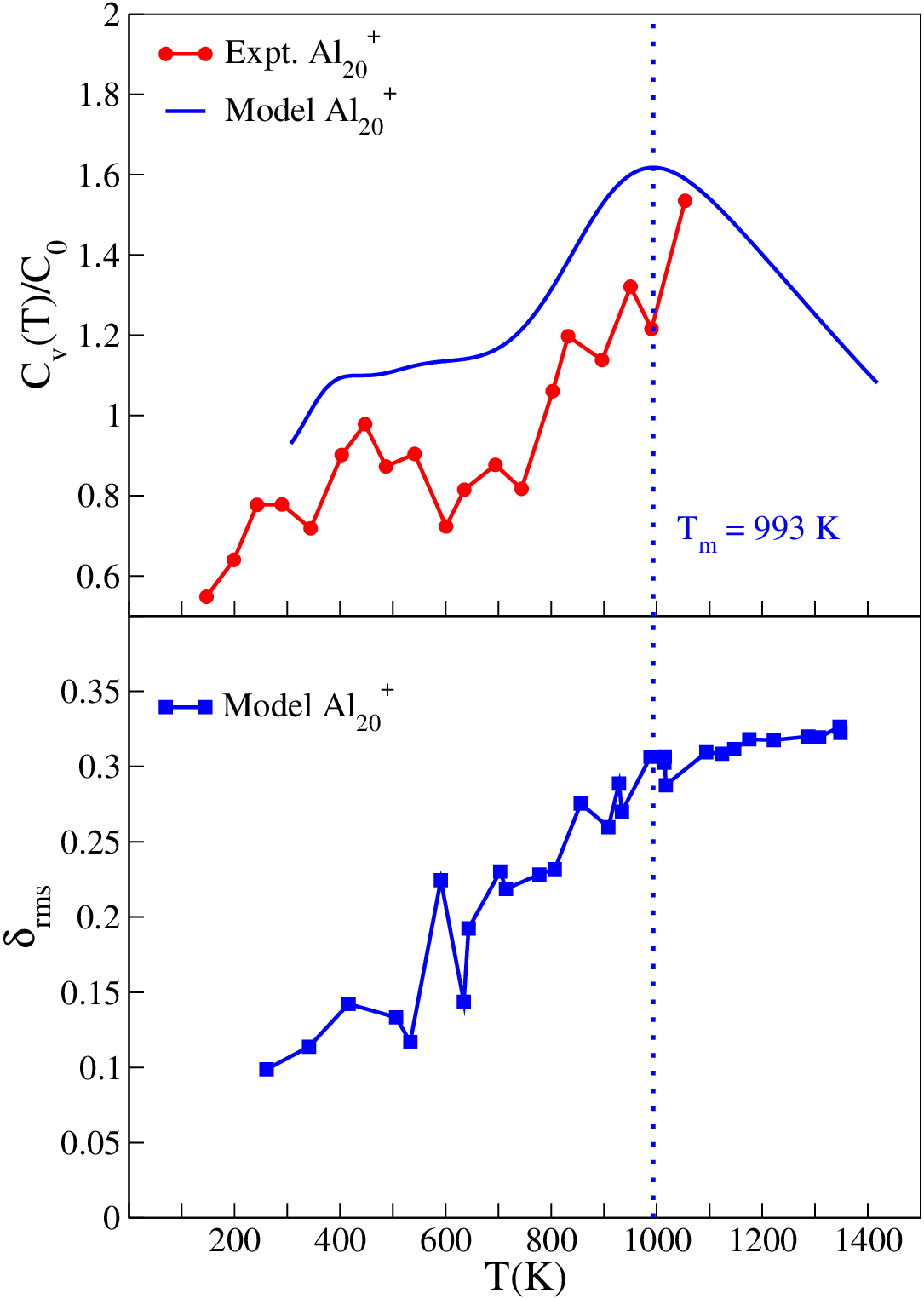}
\caption{Comparison of our modelled and experimental canonical specific heat curves (top panel) and the corresponding root mean squared fluctuations in bond-length (bottom panel) of Al\textsubscript{20}\textsuperscript{+} as a function of temperature.}
\label{spheat}
\end{figure}

Shown in the top panel of Fig.~\ref{spheat} is the comparison of experimental and our calculated canonical specific heat capacity curves normalized to the classical specific heat accounting for rotational and vibrational degrees of freedom ($C_{0} = (3N - 6 + 3/2) k_{B}$). The 29 different trajectories, spanning over the temperature range of interest, yielded evenly spaced energy histograms which were further analysed by the MH method for thermodynamics in both microcanonical and canonical ensembles.    

We report the solid-liquid-like phase transition temperature, corresponding to the maxima in specific heat curve shown in Fig.~\ref{spheat}, for Al\textsubscript{20}\textsuperscript{+} to be 993 K, much above the melting temperature of bulk aluminium (933.5 K)\cite{nist}. A comparison with the experimental canonical specific heat capacity curve\cite{Al16-48} shows that there is a strong possibility for the melting temperature to be higher than the maximum temperature spanned during experiments for Al\textsubscript{20}\textsuperscript{+} clusters.  

Although, DFT has been able to capture the melting behaviour to a high degree of accuracy in different systems, the type of functionals used also tend to have an effect on the final melting temperature. Comparison with the observed shift in melting-like transitions for charged gallium clusters\cite{Krista_PRB} would hint towards a melting temperature of ~1080 K; i.e. it seems plausible that Al\textsubscript{20}\textsuperscript{+} clusters may show melting-like features at a higher temperature if explored experimentally. 

Fluctuations in bond-length also serves, empirically, as a tool to characterize the solid-liquid-like phase transition in systems of finite sizes. The root mean squared (RMS) bond-length fluctuation criteria (often referred as the Berry parameter\cite{Berry_parameter} or Lindemann-like index) tries to capture the onset of melting as the temperature above which the average change in bond-length becomes roughly constant with increasing temperature. Calculated as $\delta_{rms}$,
\begin{equation}
  \delta_{rms} = \frac{2}{N\cdot (N-1)}\sum_{i>j}\frac{(<r_{ij}^{2}>_{t} - <r_{ij}>_{t}^{2})^{1/2}}{<r_{ij}>t}
\label{pdfeqn}
\end{equation}
where N being the total number of atoms, $r_{ij}$ being the bond distance between the participating atoms i and j and $<>_{t}$ denoting the time average, it is the bond vibrations which are quantified using this parameter. With increasing temperature the fluctuations in bond-length increase in the solid phase. However, in the liquid phase these fluctuations seems to assume a value (0.1-0.15 in bulk) which is not sensitive to temperature.  

As can be seen from the calculated RMS bond-length fluctuations from the bottom panel in Fig. \ref{spheat}, the $\delta_{rms}$ value is actually smeared over a range of temperatures (due to finite size effects), as opposed to exhibiting a sharp increase at the melting temperature . However, the values converge above the calculated melting temperature of 993 K, supporting the conclusion that there indeed is a phase transition. 

We note that although, in this case, the low and high temperature $\delta_{rms}$ values are consistent with the parallel tempering (PT) independent simulations, the absolute value of the index in the temperature range close to melting can be affected. Alternative measures unaffected by parallel tempering have also been investigated for signatures of phase transition in these Al\textsubscript{20}\textsuperscript{+} clusters.  

\subsection{Cluster geometry and structural dynamics}
\begin{figure}[t]
\subfigure[~Stacked plane]{\includegraphics[trim = 0cm 0cm 0cm 0cm,clip=true,width=0.495\columnwidth]{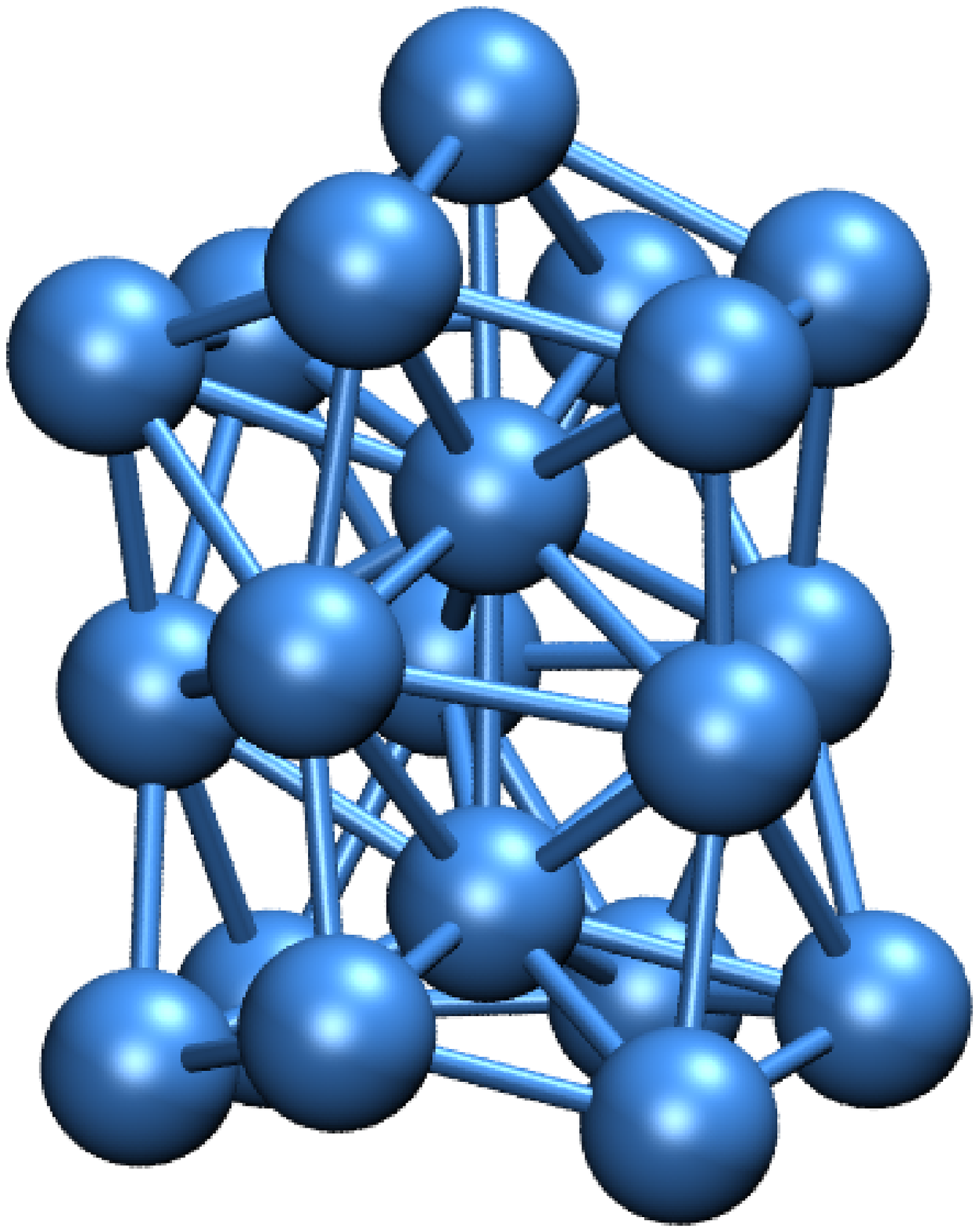}}
\subfigure[1-5-1-5-1-6-1]{\includegraphics[trim = 0cm 0cm 0cm 0cm,clip=true,width=0.495\columnwidth]{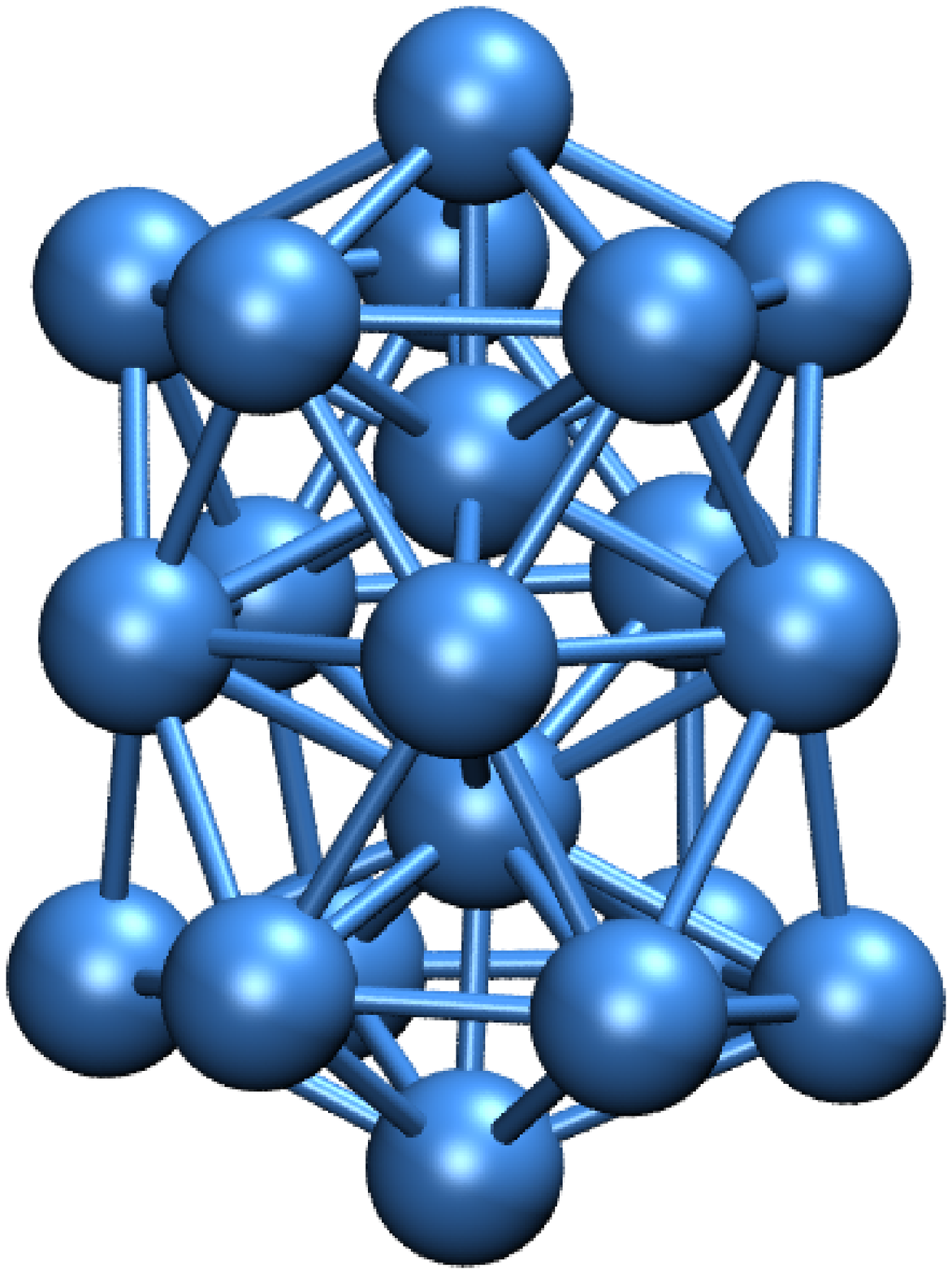}}
\caption{ Starting structure (a) and the global minimum structure (b) for Al\textsubscript{20}\textsuperscript{+}. The binding energies for SP and 1-5-1-5-1-6-1 structure being -2.796 and -2.802 eV/atom respectively.}
\label{structure}
\end{figure}
Implementation of parallel tempering MD is an efficient means of enhancing the ergodicity of the microcanonical simulations. A structure initially trapped in a local minimum of the PES would lead to incomplete sampling of the PES and hence unreliable statistics. Moreover, not only does isomerization plays a key role in the observed thermodynamics, the heat capacity also depends on the starting structures. The sharpness of the canonical specific heat capacity peak characterizes the underlying latent heat involved during the phase transition which in turn is related to the stability of the structure. For example, the icosahedron and double icosahedron Al\textsubscript{13} and Al\textsubscript{19} clusters have been seen to melt with a very sharp specific heat peak among aluminium clusters with sizes between 11 to 20 atoms\cite{Al_china}.

Our starting structure, the stacked plane (SP) configuration can also be described as 1-6-1-5-1-6 configuration where the alternate 6-atom hexagonal and 5-atom pentagonal planes are separated by a single atom as shown in Fig.~\ref{structure}(a). The SP configuration was also found to be the global minimum structure for the corresponding isoelectronic Ga\textsubscript{20}\textsuperscript{+} clusters\cite{krista1}.  

During the course of our simulations, we obtained another structure (referred to as 1-5-1-5-1-6-1 structure)\cite{Alstructure}, reported to be the global minimum structure for Al\textsubscript{20}, which is 0.006 eV lower in binding energy to the SP configuration. This structure is similar to the double icosahedron of Al\textsubscript{19} but with an embedded extra atom, thereby giving it a characteristic prolate geometry with internal structure having an impression of a mixed icosahedral stacking and hexagonal pyramid as shown in Fig.~\ref{structure}(b). Thus, although we start with a structure which is not the global minimum structure, we do find it during our MD simulation runs, thus, showing the applicability of our model in these situations.
\begin{figure}
\centering
\includegraphics[trim = 1cm 1cm 0cm 19cm,clip=true,width=\columnwidth]{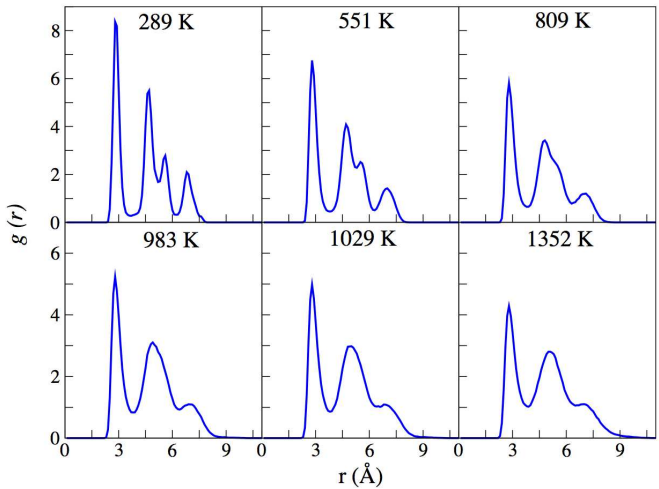}
\caption{Changes in pair distribution function (PDF) of Al\textsubscript{20}\textsuperscript{+} with average temperature during different microcanonical runs.}
\label{pdf}
\end{figure}

An analysis of the pair distribution function of Al\textsubscript{20}\textsuperscript{+}, calculated as in Eq. \ref{pdfeqn},
\begin{equation}
  g( r ) = \frac{2}{N\cdot(N-1)} \sum_{i=1}^{N} \sum_{j>i} <\delta(|r_{ij}| - r)>
\label{pdfeqn}
\end{equation}
where N is the total number of atoms in the cluster and $r_{ij}$ is the distance between atoms i and j, is as shown in Fig.~\ref{pdf}. At 289 K, the lowest average temperature obtained in our simulations, four distinct peaks are visible with the nearest neighbour distance (maxima in the plot) being 2.8 \AA. With increasing energy (corresponding to increasing average temperatures) in the subsequent microcanonical energies, the second and third peaks start to merge and beyond 809 K, there are just three peaks obtained.   

\subsection{Velocity auto-correlation function and power spectrum}
\begin{figure}
\includegraphics[trim = 0cm 0cm 0cm 0.4cm,clip=true,width=\columnwidth]{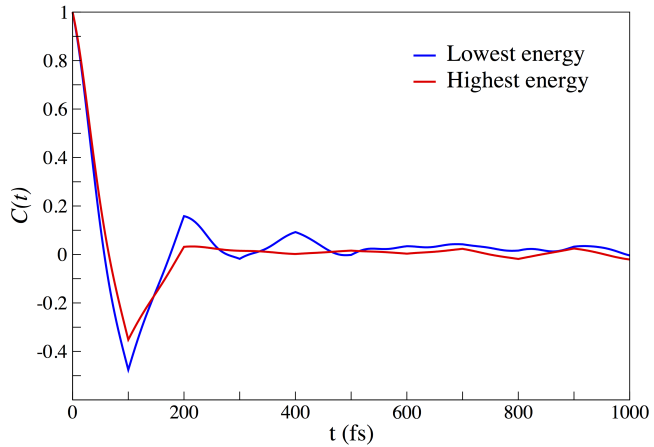}
\caption{Velocity auto-correlation function of all atoms of Al\textsubscript{20}\textsuperscript{+} cluster averaged at the lowest (blue) and highest energy (red) as a function of time.}
\label{vacfgraph}
\end{figure}

Unaffected by the parallel tempering swaps, the velocity auto-correlation function (VACF) and corresponding power spectrum provides a useful way to distinguish the solid-like phase from the corresponding liquid-like phase of a cluster. Calculated as per Eq.~\ref{vacf},
\begin{equation}
  C(t) = \frac{\sum_{i=1}^{N}\sum_{j=1}^{M}\vec{v_{i}}(t_{0j})\cdot \vec{v_{i}}(t_{0j} + t)}{\sum_{i=1}^{N}\sum_{j=1}^{M}\vec{v_{i}}(t_{0j})\cdot \vec{v_{i}}(t_{0j})}
\label{vacf}
\end{equation}
it measures the correlations in atomic motion with time. The solid-like phase of a cluster is characterised by a strongly correlated atomic motion, thus oscillatory behaviour of $C(t)$ is to be expected with time. The correlations in velocity tend to disappear with increasing energy and hence, for a liquid state $C(t)$ is relatively flat and close to zero. Fig.~\ref{vacfgraph} shows the VACF for the lowest and highest energy simulation runs. Oscillatory behaviour of $C(t)$ for the lowest energy Al\textsubscript{20}\textsuperscript{+} cluster, shown in blue, indicates strong correlations in atomic motion compared to the highest energy case, shown in red, where the curve meanders around zero indicating no correlation. This further shows that there is a solid-liquid-like phase transition occurring in these systems.     

\begin{figure}[t]
\includegraphics[trim = 0cm 4cm 0cm 2cm,clip=true,width=\columnwidth]{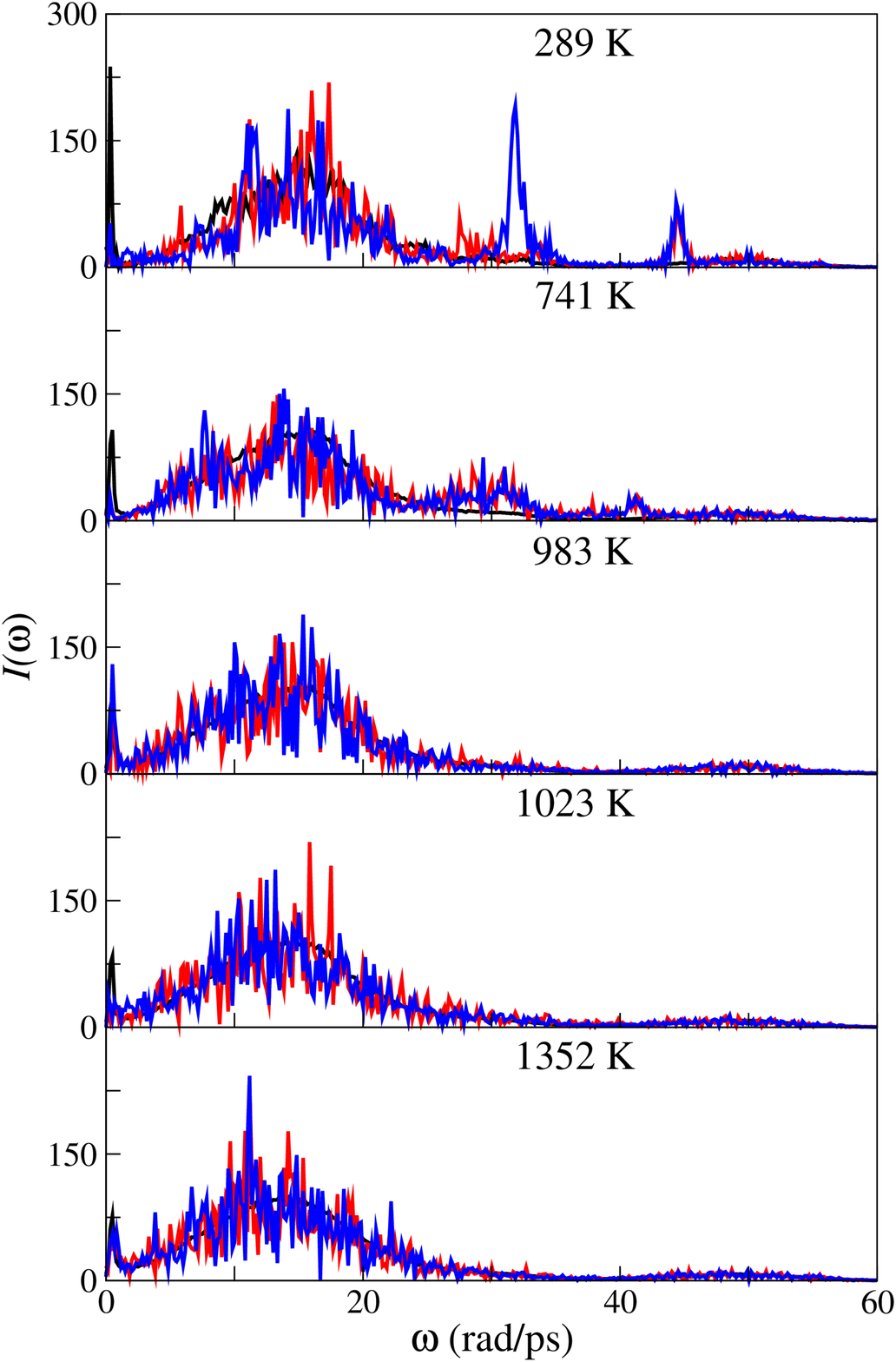}
\caption{The power spectrum of the two internal atoms (blue and red) along with that averaged for all the surface atoms (black) of Al\textsubscript{20}\textsuperscript{+} cluster as a function of frequency.}
\label{powerspec}
\end{figure}

The power spectrum or vibrational spectrum, calculated as per Eq.~\ref{pspec}, is the Fourier transform of VACF $C(t)$ with $\omega$ being the cyclic frequency. The obtained spectrum is the observation of correlations in atomic motion as a function of frequency. Moreover, individual atomic $I^{i}(\omega)$ obtained from $C^{i}(t)$ can help to discern information about the individual atomic processes. 
\begin{equation}
  I(\omega) = 2 \int_{0}^{\infty} C(t)\cdot cos(\omega t) dt
\label{pspec}
\end{equation}
 
Power spectra can also help in identifying the transition from a solid-like phase to a liquid-like phase. The criteria, as set by Yen \textit{et al.}\cite{yen}, for the phase transition temperature is the temperature at which the power spectra of internal and surface atoms overlap, also the low frequency $\omega$ becoming indistinguishable for all the atoms. Fig.~\ref{powerspec} shows the power spectra of averaged surface atoms, black, with the two internal sites, red and blue, at five different average temperatures. As the temperature increases, the high frequencies of the internal atoms starts to dissolve to that of the surface atoms and it is between 983 K and 1023 K that we observe a transition from a near-overlap to a complete overlap. The melting temperature, as obtained from the peak in the canonical specific heat curve being 993 K agrees with the observed behaviour in power spectra.   

\section{Discussion}
Higher-than-bulk melting temperatures have already been observed in experiments on gallium clusters\cite{jarrold_GaPRL} and subsequently captured in theory\cite{krista1, kristaPRB}. The differences and similarities between the clusters of these isoelectronic group 13 elements are instructive in developing a better understanding of how the nature of bonding (whether covalent or metallic) relates to the melting behaviour: in gallium clusters, claimed signatures of covalency seem to rely on an over interpretation of the Electron Localisation Function\cite{kristaPCCP,Kanhere2}. Such interpretations are completely non trivial, as a chemical `bond' is not a quantum mechanical observable, and interpretation of the nature of bonding require some understanding of the orbitals involved\cite{Ralf}.
\begin{figure}
\subfigure[~Stacked plane]{\includegraphics[trim = 0cm 0cm 0cm 0cm,clip=true,width=0.495\columnwidth]{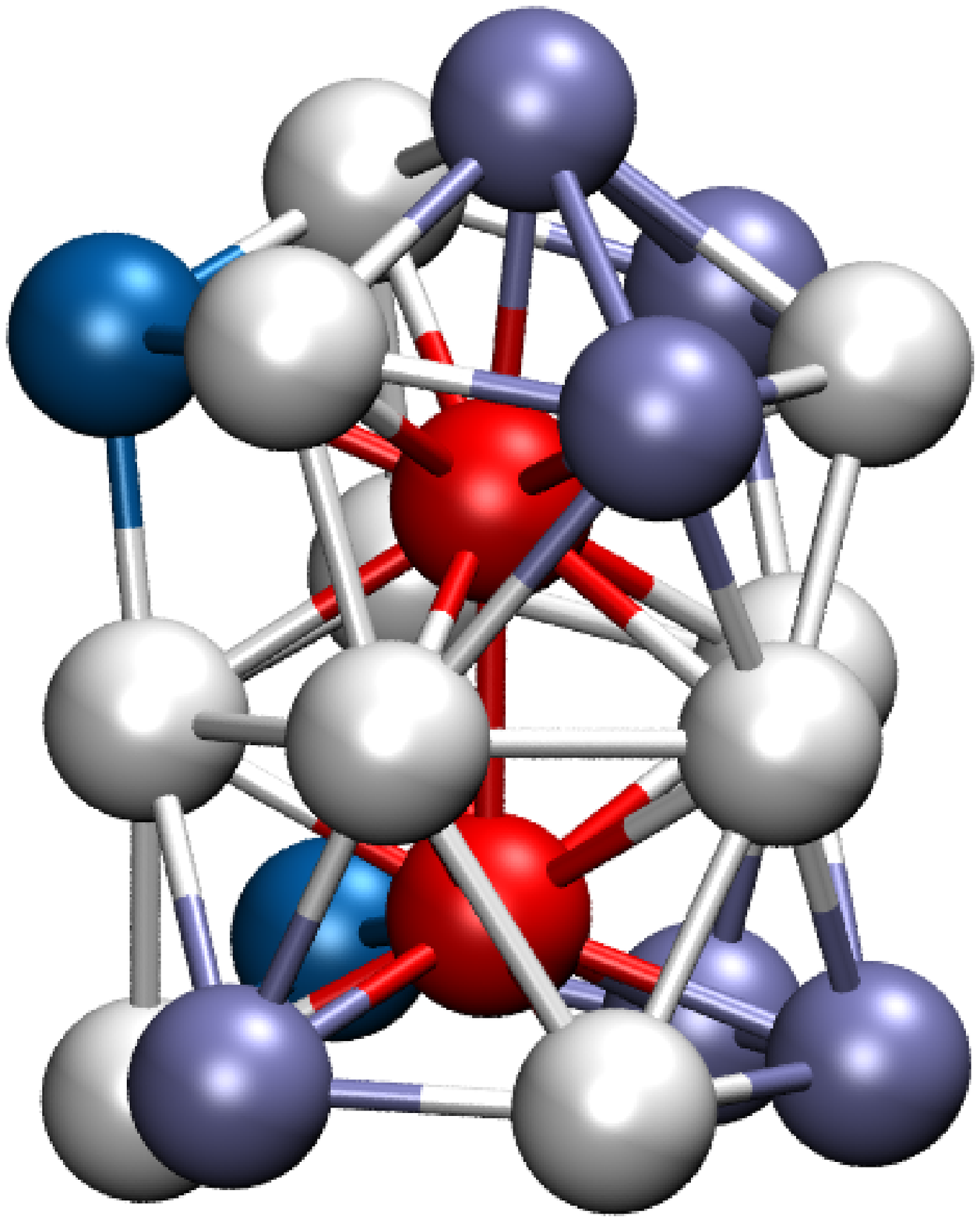}}
\subfigure[1-5-1-5-1-6-1]{\includegraphics[trim = 0cm 0cm 0cm 0cm,clip=true,width=0.495\columnwidth]{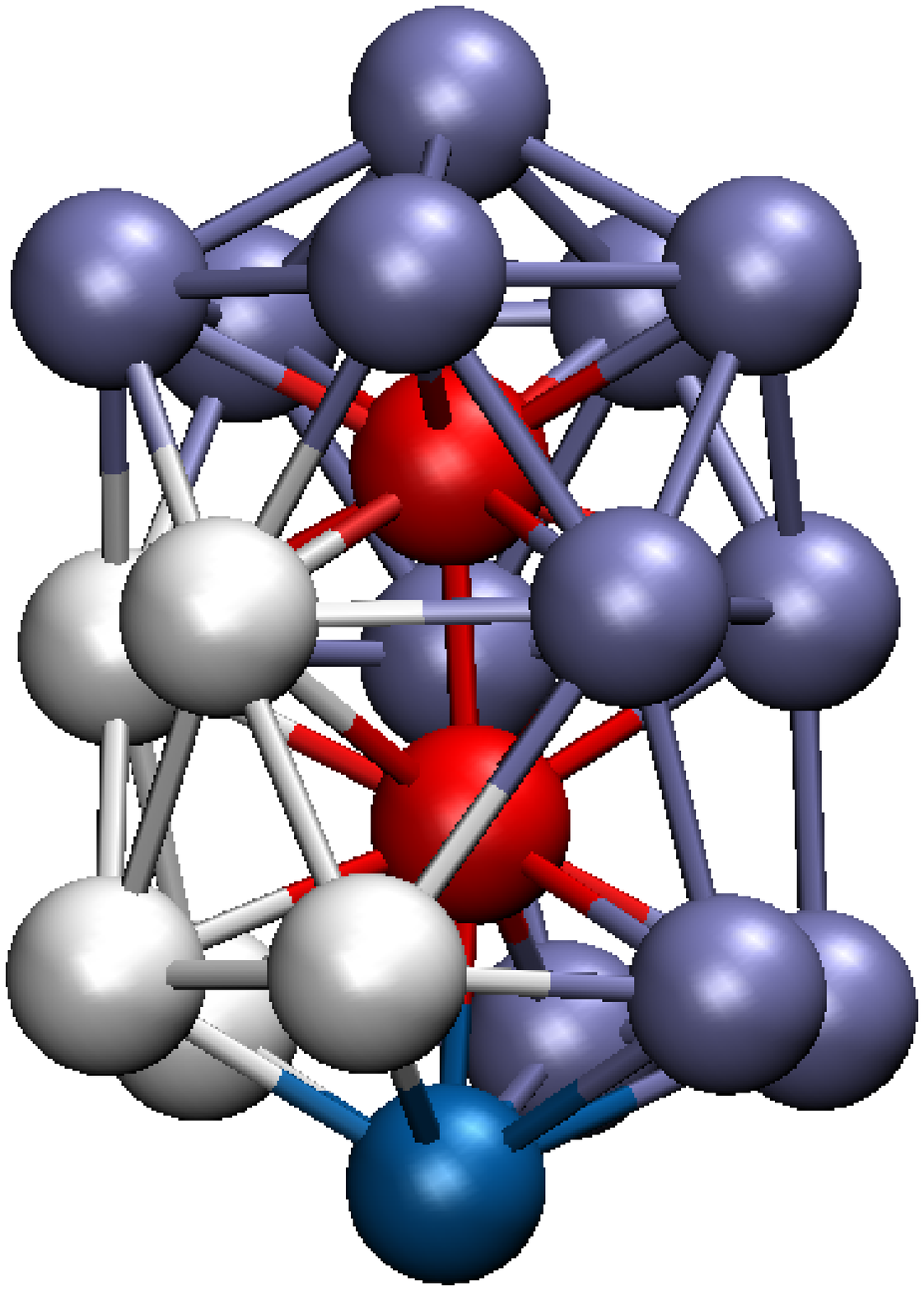}}
\caption{ Partial charges (q) obtained by atoms in molecules (AIM) analysis for Al\textsubscript{20}\textsuperscript{+} clusters: (red) q $\leq$ -0.2e ; (pink) -0.2e $<$ q $\leq$ -0.1e ; (white) -0.1e $<$ q $\leq$ 0.1e ; (ice blue) 0.1e $<$ q $\leq$ 0.2e ; (dark blue) q $>$ 0.2e.}
\label{bader}
\end{figure}
\begin{figure}[t]
\centering
\includegraphics[trim = 0cm 0cm 0cm 0cm,clip=true,width=\columnwidth]{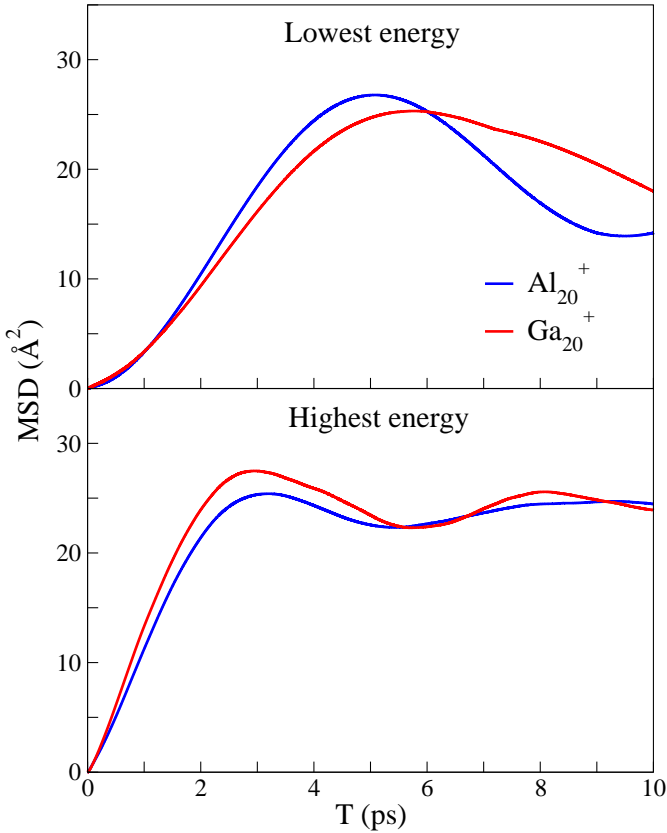}
\caption{MSD comparison of the surface sites at the lowest and highest simulation energy for Al\textsubscript{20}\textsuperscript{+} (shown in blue) and Ga\textsubscript{20}\textsuperscript{+} (shown in red) respectively.}
\label{msd}
\end{figure}

\begin{figure}[t]
\centering
\includegraphics[trim = 0cm 0cm 0cm 0cm,clip=true,width=\columnwidth]{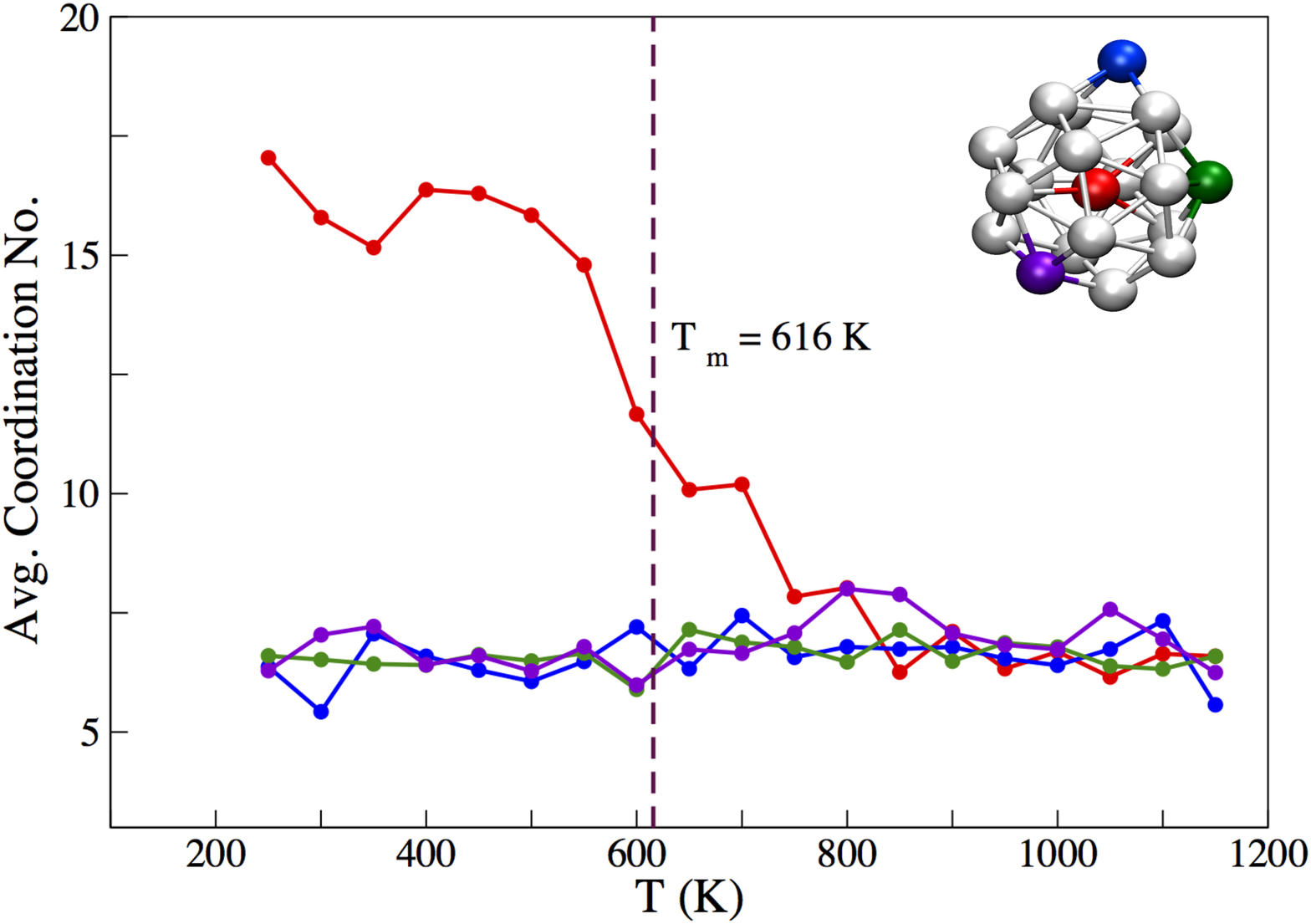}
\caption{Average coordination number variations with temperature for gallium atoms in the capped sphere (CS) configuration of Ga\textsubscript{20}\textsuperscript{+} cluster for atoms at positions shown in inset.}
\label{Ga20acn}
\end{figure}

The experimentally observed solid-liquid-like melting temperature for Ga\textsubscript{20}\textsuperscript{+} is 705 K\cite{kanherejarrold} and DFT calculations (under)estimate it at 616 K\cite{krista1}. The SP configuration, which was employed in this work, has been found to be the putative global minimum structure for Ga\textsubscript{20}\textsuperscript{+} clusters, while another class of structures, referred to as the capped sphere (CS) geometry (a single atom is caged by nineteen surface atoms with one of them giving an impression of a protruding cap), was found to be the most stable structure at finite temperatures (as also shown in the inset of Fig.~\ref{Ga20acn}). The CS structure, however, is not found for Al\textsubscript{20}\textsuperscript{+} clusters. 

The Quantum Theory of Atoms-In-Molecules (QTAIM), as proposed by Bader\cite{bader}, provides a useful way to associate partial charges with each participating ion in the overall unit positively charged gallium and aluminium clusters. Atomic boundaries within a molecule is defined based on the topology of the total electronic charge density of the molecular system.  An integration of the electron density within each atomic volume gives the partial charge of the corresponding ion. Fig.~\ref{bader} shows the partial charges obtained for Al\textsubscript{20}\textsuperscript{+} in, both, the SP and the putative global minimum structure (1-5-1-5-1-6-1). There is strong charge segregation between the internal and surface sites with the internal site becoming strongly negatively charged and the surface atoms becoming positive. This kind of charge distribution within the cluster was also observed in Ga\textsubscript{20}\textsuperscript{+} in its putative global minimum SP geometry. However, it was only the internal and the capping atom that were negatively and positively charged respectively in the CS structure\cite{krista1}. Negative charge at internal sites, and the surface sites becoming positively charged seems to be a common feature in both Al\textsubscript{20}\textsuperscript{+} and Ga\textsubscript{20}\textsuperscript{+} clusters.

\begin{figure}[t]
\centering
\includegraphics[trim = 0cm 0cm 0cm 0cm,clip=true,width=\columnwidth]{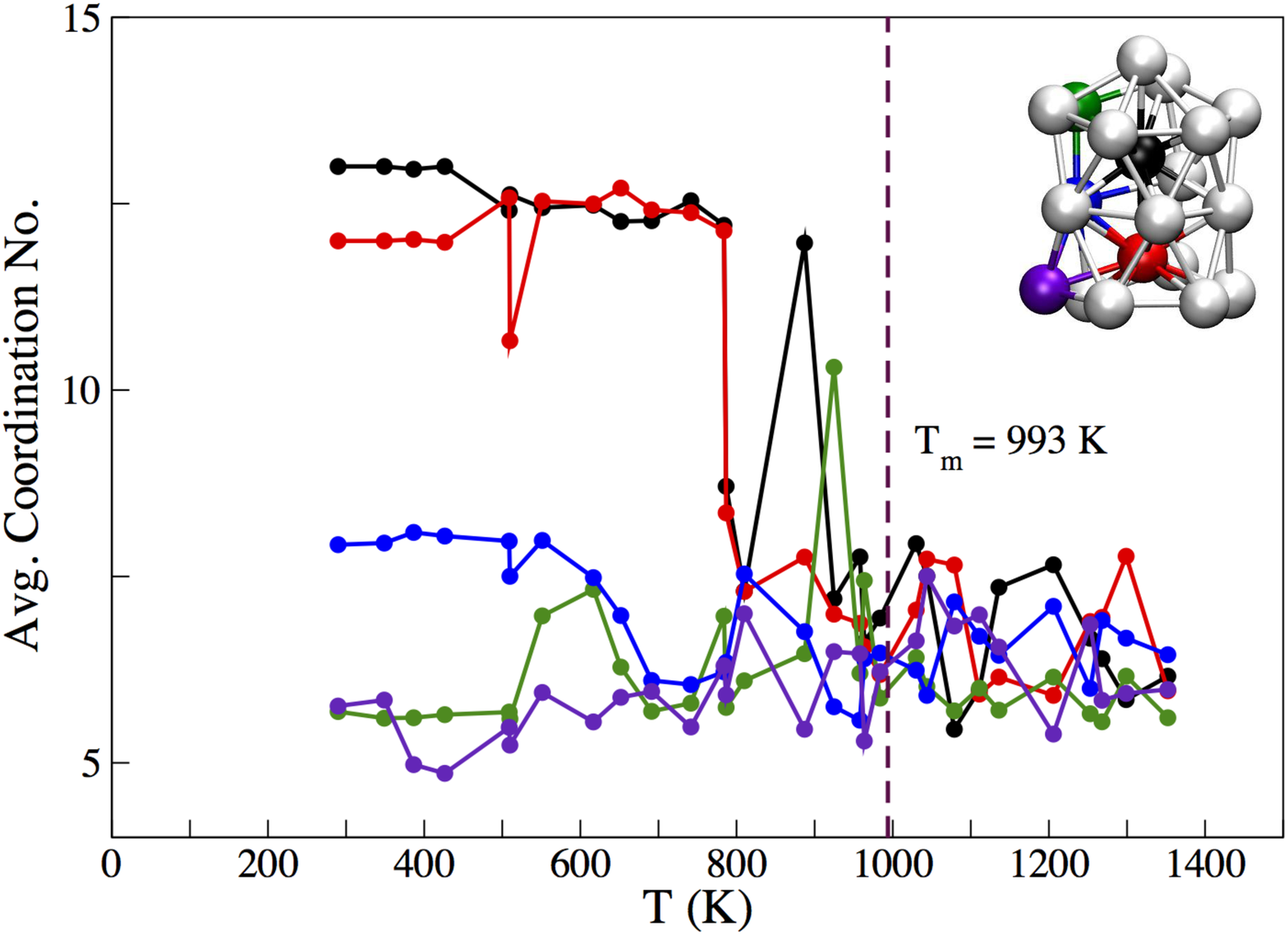}
\caption{Average coordination number variations with temperature for aluminium atoms at positions shown in inset.}
\label{can}
\end{figure}

The charge segregation may affect the relative mobility of the internal and surface sites of Al\textsubscript{20}\textsuperscript{+} and Ga\textsubscript{20}\textsuperscript{+} clusters. Analysis of the mean squared displacements provides further insight into how the motion of atoms situated at different positions, i.e. internal or surface, affects the observed kinetics and thermodynamics of the cluster. It is calculated as per Eq.~\ref{msdeqn},
\begin{equation}
  <r_{i}(t)> = \frac{1}{M}\sum_{m=1}^{M}[R_{i}(t_{m}+t) - R_{i}(t_{m})]^{2}
\label{msdeqn}
\end{equation}
where \textit{M} is the number of time origins and \textit{i} is the atom number. Shown in Fig.~\ref{msd} is the averaged mean squared displacement of the surface atoms at the lowest and highest simulated energy for Al\textsubscript{20}\textsuperscript{+} and Ga\textsubscript{20}\textsuperscript{+} clusters. The mobilities of the surface atoms are relatively similar at the lowest and highest energies for both the clusters indicating that the increase in energy has a very small effect on the mobility of surface atoms.  

An assessment of the structural changes occurring in Al\textsubscript{20}\textsuperscript{+} cluster with temperature can be discerned from the average coordination number of each participating atom. Fig.~\ref{can} shows the average coordination number for selected atoms as coloured in inset. There are three distinct coordination sites at low temperatures: internal atoms (black and red), surface atoms situated in the central ring (blue) and the top and bottom ring surface atoms (green and violet). However, after the melting temperature, all the atoms have nearly equal coordination numbers on average. In contrast, the average coordination number of Ga\textsubscript{20}\textsuperscript{+} cluster, as shown in Fig.~\ref{Ga20acn}, shows the presence of only two distinct sites, viz. the central site (red) and the surface sites (three representative surface atoms coloured in green, blue and violet).      


In order to probe further the motion of atoms in Al\textsubscript{20}\textsuperscript{+} and Ga\textsubscript{20}\textsuperscript{+} clusters, we calculated the self-diffusion coefficients for both cases. At sufficiently long times, the mean squared displacement of an atom is proportional to the observation time. The self-diffusion coefficient, expressed as D, is calculated as per Eq.~\ref{selfdiffcoeff},
\begin{equation}
D \equiv \frac{1}{2d}\lim_{t\rightarrow\infty}\frac{<r_{i}(t)>}{t}
\label{selfdiffcoeff}
\end{equation}
where the numerator represents the mean squared displacement as described in Eq.~\ref{msdeqn} and $d$ is the dimensionality of the system. A linear least squares regression fit to the obtained MSD data is performed to calculate the slope, which is equal to $(2dD)$, from which the self diffusion coefficient is calculated. It must be noted that for finite systems such as clusters, the self diffusion coefficient carries a meaning only over a small time scale where the slope is constant\cite{berry_msd}. 

\begin{figure}[t]
\centering
\includegraphics[trim = 0cm 0cm 0cm 0cm,clip=true,width=\columnwidth]{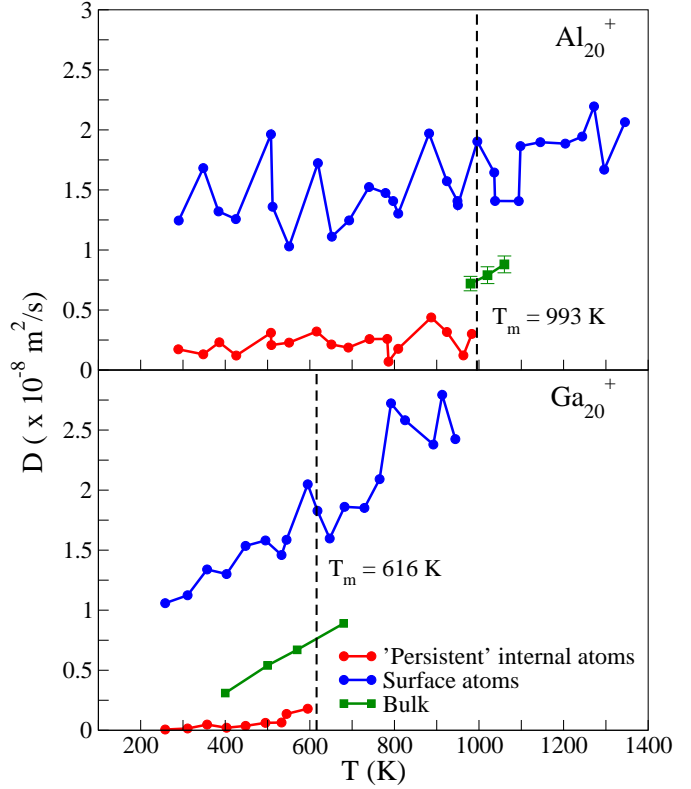}
\caption{Comparison of self diffusion coefficient of internal and surface atoms Al\textsubscript{20}\textsuperscript{+} (top panel) and Ga\textsubscript{20}\textsuperscript{+} (bottom panel) along with the corresponding value in the bulk liquid phase of aluminium\cite{Al_selfdiff} and gallium\cite{Ga_selfdiff} as a function of temperature.}
\label{selfdiff}
\end{figure}

The average coordination number of the atoms in the Al\textsubscript{20}\textsuperscript{+} cluster, Fig.~\ref{can}, shows that at 924 K i.e. below the melting temperature, one of the atoms in the top ring (shaded green) has an average coordination number typical for an internal site. This is contrasted with the atom (shaded red) which was at an internal site at the lowest temperature. Atoms of the Al\textsubscript{20}\textsuperscript{+} cluster are seen to swap positions often near the melting temperature as also confirmed from the MD movies, making it much harder to identify and tag a particular atom as internal. However, this is not observed in the Ga\textsubscript{20}\textsuperscript{+} cluster. To counter this situation, we attribute the term `persistent' internal to an atom which satisfies the following two conditions: (a) the average coordination number has to be greater-than-or-equal-to 9.5 and, (b) the time during which condition (a) is satisfied during the MD run has to be more than 1.5 ps. 

Shown in Fig.~\ref{selfdiff} is the self-diffusion coefficient (D) calculated for Al\textsubscript{20}\textsuperscript{+} and Ga\textsubscript{20}\textsuperscript{+} in the top and bottom panel respectively. The MSD values for each atom over which the self-diffusion coefficients have been calculated is between 1 and 2 \AA\textsuperscript{2}. A striking similarity below and after the melting temperatures is observed in both cases. The surface atoms of both, Al\textsubscript{20}\textsuperscript{+} and Ga\textsubscript{20}\textsuperscript{+}, clusters show very high self-diffusion coefficient with the trend increasing with temperature. The magnitude of self-diffusion coefficient for the surface atoms of Al\textsubscript{20}\textsuperscript{+} is comparable to that of bulk aluminium in liquid phase\cite{Al_selfdiff}. However, it is the `persistent' internal atoms whose self-diffusion coefficient is very low in comparison to the corresponding surface atoms below the respective melting temperatures.  After the melting temperature, we do not find any of the 20 participating atoms to satisfy the `persistent' internal atom criterion. Individual atomic self-diffusion coefficients, not shown here, after the melting temperature have nearly the same self-diffusion coefficients as the surface atoms, reflecting a liquid-like phase. This demonstrates that it is the internal atoms which are responsible for the melting-like feature, thereby contributing to the latent heat as shown in the specific heat capacities (top panel of Fig.~\ref{spheat}). The strong charge segregation between the internal and surface atoms, with negative charging of the internal atoms, indicates that the interaction between the internal and surface atoms is largely electrostatic. The significance of this point is reflected in the significantly lower self-diffusion coefficient of the internal atoms in the clusters than for the atoms in the respective bulk materials: the internal atoms in these clusters are not only distinct from the surface atoms, they cannot be considered to be intermediate between the surface and the bulk, either. 
Given that the melting transition corresponds to the movement of the internal atom into the liquid state (same mobility as the surface atoms), the nature of this electrostatic interaction should be considered critical for the melting transition. 

\section{Conclusions}
First-principles calculations predict that small aluminium clusters, in particular Al\textsubscript{20}\textsuperscript{+}, may exhibit a melting transition at temperatures above the bulk melting temperature. This suggests a enhanced similarity between gallium and aluminium clusters at these small sizes, which is lost at larger sizes. Analysis of the environment of the internal atoms in these clusters indicates the importance of an electrostatic cage in confining the internal atoms, the disruption of which is required for melting to occur.  This picture of melting in these clusters, which emphasises the role of the surface, and suggests a natural limit to greater-than-bulk melting temperatures at sizes where the internal atoms occupy an environment closer to that of bulk atoms, should be considered as a useful step towards a physical description of the causes of greater-than-bulk melting temperatures in clusters.

\section{Acknowledgement}
The authors would like to acknowledge the support of Royal Society of New Zealand through the Marsden Fund under contract no. IRL0801 and New Zealand eScience Infrastructure (NeSI) (Project ID NESI97) for computational resources. 

\bibliography{bibliography_arxiv}
\bibliographystyle{plain}



\end{document}